\documentclass[12pt,preprint] {aastex}

\usepackage{epsfig}
\usepackage{psfig}
\usepackage{natbib}

\DeclareMathAlphabet{\mathsc}{OT1}{cmr}{m}{sc}
\def\testbx{bx}%
\DeclareRobustCommand{\ion}[2]{%
\relax\ifmmode
\ifx\testbx\f@series
{\mathbf{#1\,\mathsc{#2}}}\else
{\mathrm{#1\,\mathsc{#2}}}\fi
\else\textup{#1\,{\mdseries\textsc{#2}}}%
\fi}
\def\squareforqed{\hbox{\rlap{$\sqcap$}$\sqcup$}}
\def\sq{\ifmmode\squareforqed\else{\unskip\nobreak\hfil
\penalty50\hskip1em\null\nobreak\hfil\squareforqed
\parfillskip=0pt\finalhyphendemerits=0\endgraf}\fi}
\def\arcmin{\hbox{$^\prime$}}
\def\arcsec{\hbox{$^{\prime\prime}$}}
\def\diameter{{\ifmmode\mathchoice
{\ooalign{\hfil\hbox{$\displaystyle/$}\hfil\crcr
{\hbox{$\displaystyle\mathchar"20D$}}}}
{\ooalign{\hfil\hbox{$\textstyle/$}\hfil\crcr
{\hbox{$\textstyle\mathchar"20D$}}}}
{\ooalign{\hfil\hbox{$\scriptstyle/$}\hfil\crcr
{\hbox{$\scriptstyle\mathchar"20D$}}}}
{\ooalign{\hfil\hbox{$\scriptscriptstyle/$}\hfil\crcr
{\hbox{$\scriptscriptstyle\mathchar"20D$}}}}
\else{\ooalign{\hfil/\hfil\crcr\mathhexbox20D}}%
\fi}}
\begin{document}
 
\title {On the origin of the \ion{H}{i} holes in the interstellar medium of dwarf irregular galaxies}

\altaffiltext {1}{Max-Planck-Institut f\"{u}r Astronomie K\"{o}nigstuhl 17, D-69117 Heidelberg, Germany; dib@mpia.de}
\altaffiltext {2}{University Observatory Munich, Scheinerstrasse 1, D-81679 Munich, Germany; burkert@usm.uni-muenchen.de }
\altaffiltext {3}{On leave to Centro de Astronom\'{i}a y Astrof\'{i}sica, UNAM, Apdo. 72-3 (Xangari), 58089 Morelia, Michoac\'{a}n , Mexico}

\author{Sami Dib\altaffilmark{1,2,3}, Andreas Burkert\altaffilmark{2}}

\begin{abstract}
We suggest that large \ion{H}{i} holes observed in the interstellar medium (ISM) of galaxies such as the Large Magellanic Cloud (LMC), the Small Magellanic Cloud (SMC) and \objectname{Holmberg II} (Ho II, DDO 50, UGC 4305) can form as the combined result of turbulence coupled with thermal and gravitational instabilities. We investigate this problem with three dimensional hydrodynamical simulations, taking into account cooling and heating processes and the action of the self-gravity of the gas. We construct an algorithm for radiative transfer to post-process the simulated data and build emission maps in the 21 cm neutral hydrogen line. With this approach, we are able to reproduce the structure of the shells and holes as observed in regions of the ISM, where no stellar activity is detected. In order to quantify the comparison of our synthetic maps to the observations, we calculate the physical scale-autocorrelation length relation ($L-L_{cr}$ relation) both on the synthetic \ion{H}{i} maps and the \ion{H}{i} map of Ho II. The $L-L_{cr}$ relation shows a linear increase of the autocorrelation length with the physical scale up to the scale of energy injection and flattens for larger scales. The comparison of the $L-L_{cr}$ relation between the observations and the synthetic maps suggests that turbulence is driven in the ISM of Ho II on large scales ($\sim~6$ kpc). The slope of the $L-L_{cr}$ relation in the linear regime in Ho II is better reproduced by models where turbulence is coupled  with a weak efficiency cooling of the gas. These results demonstrate the importance of the interplay between turbulence and the thermodynamics of the gas for structure formation in the ISM. Our analysis can be used to determine the scale on which kinetic energy is injected in the ISM of dwarf irregular galaxies, and to derive, in a first approximation, the cooling rate of the gas.
\end{abstract} 

\keywords{instabilities --- turbulence --- ISM: atoms --- galaxies: ISM --- ISM: structure --- methods: numerical}

\section{INTRODUCTION}

The question of which physical processes produce the holes observed in the 21 cm \ion{H}{i} line in the interstellar medium (ISM) of many galaxies such as in the Large Magellanic Cloud (LMC) (Kim et al 1999), Holmberg II (Ho II, DDO 50, UGC 4305) (Puche et al. 1992), the Small Magellanic Cloud (SMC) (Staveley-Smith et al. 1997, Stanimirovi\'c et al. 1999) and the Western Magellanic Bridge (Muller et al. 2003) has been a controversial and puzzling issue in the past. On the observational side, there are evidences that many of the shell and void-like structures observed in the ISM of different galaxies, especially in the outer parts of their galactic disks are unlikely to be the result of supernova explosions. Kim et al. (1999) came to this conclusion for the LMC in view of the weak correlation that exists between the positions of some of the \ion{H}{i} shells and the \ion{H}{ii} regions. Using a multicolor BVR imaging approach, a careful study by Rhode et al. (1999) of the 51 \ion{H}{i} holes in Ho II, previously selected by Puche et al. (1992), shows no signs of the presence of remnant stellar clusters inside 86 percent of these holes. Additionally, using H$\alpha$ photometry, they could not confirm the presence of any ongoing star formation process inside the holes. Based on these observations, they concluded that those holes can not be supernova driven expanding shells. Furthermore, Stewart et al. (2000) observed the far-ultraviolet (FUV) emission of Ho II. The FUV emission is sensitive to massive star formation on a timescale which is comparable to the lifetime of the largest holes in Ho II ($\sim$ 100 Myrs). Only 3 out of the 51 \ion{H}{i} holes selected by Puche et al. (1992) have associated FUV emission whereas in the rest of the holes the emission is concentrated at their edges. Furthermore, only very little FUV emission is detected outside the 2$\arcmin$-3$\arcmin$ radii inner region, indicating that star formation has played only a little role in the outer region. X-ray observations of Ho II using ROSAT PSPC by Zezas et al. (1999) and Kerp et al. (2002) have shown that most of the \ion{H}{i} holes are devoid from hot gas. However, Kerp et al. (2002) showed that a few X-ray sources (sources 2 and 4 in Fig. 1 in their article), located outside the Stellar Disk but within the \ion{H}{i} distribution of Ho II have emission that might be associated with thermal plasma radiation which is characteristic of supernova remnants. This suggests that, the supernova scenario can not be entirely ruled out in the case of those few holes.

Alternative mechanisms which could lead to the creation of the \ion{H}{i} holes in Ho II have been discussed by several authors (Rhode et al. 1999; Bureau \& Carignan 2002). Some of the proposed scenarios are related to the supernova scenario, like invoking a non standard initial mass function and modifying the holes energetics which helps to reduce the number of supernovae needed to create them. The latter scenario was pointed out by Rhode et al. (1999). It would however require a reduction of the supernova rate by a factor of $\sim 5$ in order to have a remnant stellar cluster with a luminosity that falls below their detection limit. This in turn would require that the expansion velocities in the case of the largest holes have been overestimated by a factor $\sim 3$ by Puche et al. (1992), which seems unlikely. The scenario of an extremely top-heavy IMF in Ho II is also not supported by any observations of significant variations in the IMF measured in different environments like the Milky Way or other nearby galaxies (Leitherer 1999) and thus, seems also very unlikely. \ion{H}{i} holes could also be created by the impact of High-Velocity Clouds (HVCs) on the gaseous disk. In the case of the Milky Way and other galaxies, it has been shown that some holes are associated with high velocity gas (Heiles 1984 for the Milky Way; Van der Hulst \& Sancisi 1988 for M101). In the case of Ho II, only one object has been identified by Rhode et al. (1999) as a possible HVC. This object has an estimated mass of 1.2$\times$10$^{7}$ M$_{\odot}$ and a total kinetic energy of $\approx 5 \times 10^{53}$ ergs. This amount of energy is enough to create any of the observed \ion{H}{i} holes in Ho II. Aside from being statistically not very significant, this cloud has been detected on only one pixel in the whole data set. As already pointed out by Rhode et al. (1999), the signal detected from that pixel could also be assimilated to noise under the assumption that the noise in the data set is Gaussian. Only a more sensitive study of  the \ion{H}{i} structure around Ho II can shed more light on the existence and role of HVCs. By re-processing the low resolution map of Puche et al. (1992), Bureau \& Carignan (2002) have unveiled the large scale structure of the \ion{H}{i} in Ho II to be comet-like. They interpret this as gas in Ho II experiencing ram-pressure stripping under the effect of the intra-group medium (IGM) and that this process might lead to the formation of some holes or to the enlargement of existing ones. This scenario would only work if the IGM density is $n_{IGM}\gtrsim 4.0\times 10^{-6}$ cm$^{-3}$. Bureau \& Carignan (2002) pointed out to the fact that tidal interactions on the scale of the subgroup of galaxies containing Ho II, Kar 52 and UGC 4483 could also affect the \ion{H}{i} morphology of Ho II. However, a preliminary analysis of the \ion{H}{i} structure on the subgroup scale by Bureau et al. (2004) does not support the scenario of gas being tidally stripped from the disk of Ho II in the direction of Kar 52 and vice versa. It is also inconclusive with regard to the role played by ram pressure stripping.   

 In this work, we investigate whether the \ion{H}{i} morphology could also emerge as the result of internal gas dynamics as earlier suggested by Elmegreen (1997). Our aim is to explore the role played by turbulence and by thermal (Field 1965; Field et al. 1969)  and gravitational instabilities (Kolesnik 1991) (TI and GI, respectively) on the formation of large scale structure in the ISM. We use three dimensional hydrodynamical isothermal simulations of driven turbulence to investigate if holes like those in Ho II are formed as a natural consequence of the turbulence present in the medium. We then study the effect of heating and cooling and take the self-gravity of the gas into account. In a post-process approach, we build two dimensional emission maps in the  21 cm \ion{H}{i} hyperfine structure line and compare them to observations.

TI has been traditionally considered as an important agent of structure formation in the ISM. Burkert \& Lin (2000) performed a linear perturbation analysis of TI and studied numerically under which condition the transition from the linear to the non-linear regime occurs. S\'anchez-Salcedo et al. (2002), used high resolution 1D simulations to study the behavior of TI in the presence of forced flows. Koyama \& Inutsuka (2000, 2002) have investigated with 1D and 2D simulations which include detailed cooling, heating and chemistry, the propagation of shocks in the warm neutral medium (WNM) and cold neutral medium (CNM) and the development of TI in the post-shock region. Kritsuk \& Norman (2002a,b) showed, using three dimensional simulations, that a time fluctuating background heating supports turbulence in a multiphase ISM. Piontek \& Ostriker (2004) investigated the interplay between TI and the magnetorotational instability (Balbus \& Hawley 2001). V\'azquez-Semadeni et al. (2000) and Gazol et al. (2001) used local two dimensional numerical simulations to investigate TI in the ISM on a scale similar to ours (1 kpc). They focused on the segregation of the medium into two distinct gaseous phases. A major success of their model is that they were able to reproduce a mass fraction of the unstable gas which is in good agreement with the observations (Heiles 2001), but they did not build emission maps that can be compared with observations. Wada \& Norman (1999, 2001) and Wada et al. (2000) have shown that the combined effect of TI and GI with turbulence can lead to the formation of \ion{H}{i} holes. They performed global two dimensional numerical simulations of an LMC type galactic disk and constructed from their hydrodynamical data emission maps in the 21 cm \ion{H}{i}  and the CO(1-0) lines. They transformed their two dimensional grid into a three dimensional one by assuming a constant scale height for the disk, divided in a number of grid cells. Though they used a rather sophisticated approach to solve the radiation equation in each of the newly created cells, the local densities in all cells along a given line of sight were assumed to be the same, which is a major simplification. Additionally, although a similarity in the complex network of holes and shells is observed between their simulated \ion{H}{i} map and observations of the structure in the 21 cm line of the LMC, they did not quantify the hole and shell-like structures. This contrasts with our simulations which take into account the complexity of the three dimensional structure, which arises self-consistently. The \ion{H}{i} holes formation process by TI and GI in the ISM of dwarf irregular galaxies is critically reviewed by S\'anchez-Salcedo (2001). The paper is organized as follows. In \S 2 we describe our set of simulations. The radiative transfer approach we use to construct the emission maps is described in \S 3. In \S 4 we define the observable quantity which we use to compare our models with observations. In \S 5 we present the observational data and in \S 6 we evaluate the autocorrelation length in our simulated \ion{H}{i} maps and compare it with the observations. In \S 7, we summarize our results. 

\section{THE SIMULATIONS} 

We use the ZEUS-3D code (Stone and Norman 1992a,b) to solve the equations (mass, momentum and internal energy conservation) of ideal gas dynamics. In addition, we solve the Poisson equation to account for the self-gravity of the gas in some of the simulations. The Poisson equation is solved using a FFT-solver following an algorithm by Burkert \& Bodenheimer (1993) which is derived from Press et al. (1992). The grid represents a 1 kpc$^{3}$ volume of the ISM, and the calculations are performed with a resolution of 128$^{3}$ grid cells. A polytropic equation of state with a specific heat ratio of 1 is used in isothermal simulations and 1.4 in simulations including cooling and heating. Periodic boundary conditions are imposed in all three directions. The medium is originally uniform in density with an average density $\bar{n}=0.5$ cm$^{-3}$, which is the typical value for the warm interstellar medium. 

\subsection{ISOTHERMAL SIMULATIONS}

\begin{figure}
\plotone{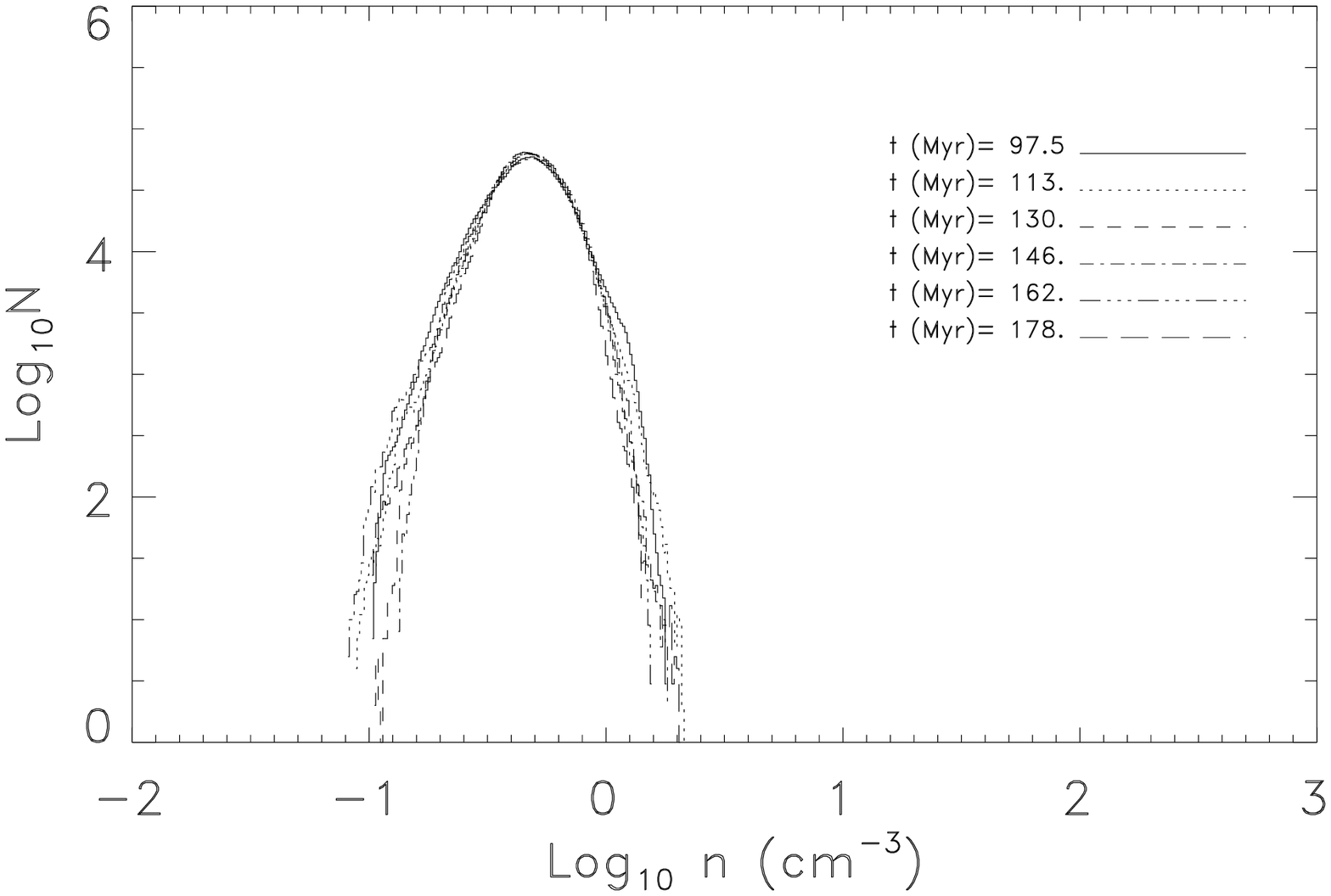}
\caption{Time evolution of the number density distribution function in simulation ISO2m1. In this plot, N is the number of cells per density bin.}
\label{fig1}
\end{figure}

We first performed a number of isothermal simulations (runs ISOimj in Tab.~\ref{tab1}) in which the turbulence is driven on different length scales and with different Mach numbers (i indicates the driving wave number and j the Mach number). The initial density fluctuations for the isothermal runs are generated following the description given in Klessen (1998) and Klessen \& Burkert (2000). The density fluctuations are described by a power spectrum $P(k)=k^{m}$. We have chosen $m=0$, which implies that the energy is distributed equally on all scales. We should however mention that since the medium is constantly driven, its evolution is rather independent of the initial conditions. We mimick the turbulence driving by compensating for the energy lost through dissipation by injecting random motions to the gas such that the total injected energy is equal to the energy-dissipation rate following the description of Mac Low (1999). With this prescription, the medium is driven at a certain scale which is defined by a certain characteristic wave number $k$ (i.e., characteristic driving scale $L_{driv}=2\pi/k$). The medium is evolved over two dynamical timescales. The dynamical timescale is defined as the ratio of the box size $L_{b}$ to the initial {\it rms} turbulent velocity $\tau_{dyn}=L_{b}/v_{rms,0}\sim10^{8}$ years, where $L_{b}$=1 kpc and $v_{rms0}=10$ km sec$^{-1}$. The value of 10 km s$^{-1}$ has been chosen as being a typical value of the warm phase in the ISM. In these isothermal simulations, the medium reaches a dynamical equilibrium (stationary probability distribution functions, PDFs, of the density, see Fig.~\ref{fig1}) after $\sim$ one dynamical timescale with the largest over-densities being of the order of $\delta\rho/\rho\sim 3$.  

\subsection{NON-ISOTHERMAL SIMULATIONS}  

The density and velocity distributions inherited after evolving the medium for one dynamical timescale in the isothermal regime (with $k=2$ and $Ma=1$) are used as initial conditions for simulations including cooling and heating (Runs CO0jGi in Tab.~\ref{tab1}). G stands for runs where self-gravity is included (T otherwise), j describes the cooling efficiency (see below) and i the above mentioned wave number value. The simulations are evolved for two dynamical times after heating and cooling are turned on. The timestep that is used to evolve the internal energy is the CFL timestep which means that the energy net losses/gains are assumed to affect implicitely the determination of the timestep at each iteration. In our simulations, this can be justified by the fact that the largest changes in the internal energy in a given cell do not exceed a few percents of the initial available energy reservoir. A similar approach has been adopted by several authors who have used the ZEUS code in astrophysical problems that involve cooling and heating of the gas (e.g., Pavlovski et al. 2002; Smith et al. 2003). An initial {\it rms} Mach number of 1 is reasonable given the observed \ion{H}{i} velocities of a few km s$^{-1}$ with temperatures of a few 1000 K. The initial density distribution for the simulations with cooling is inhomogeneous with ($\delta\rho/\rho)_{max}=2.8$ and temperature is set initially everywhere to $10^{4}$ K. We assume the cooling and heating terms in the energy equation to be, respectively, $\Lambda(\rho,T)=\Lambda_{0}\rho^{\alpha}T^{\beta}$ and $L(\rho)=q\Lambda_{0}\rho^{\epsilon}$. The later term is responsible for keeping a non-zero fraction of the gas in the intermediate temperature regime by heating up the envelopes of dense clouds and restoring cold gas to the diffuse phase. The temperature is allowed to vary between $T_{min}=10$ K and $T_{max}=12000$ K. We set $\alpha=2$ mimicking the density squared dependence in the cooling function describing radiative cooling (Spitzer 1972) and $\epsilon=1$, thus mimicking a constant background heating. We use a simple prescription to account for the self-shielding of the clouds by nulling the heating term in the inner cells of a cloud and applying a shape dependent linear heating at the boundary cells of the cloud. This introduces a shielding density, $n_{shield}$ above which the shielding of the clouds becomes effective. An inner cell is defined as a cell where all of its 26 neighoring cells have a density which is higher then the shielding density. A surface cell is defined as a cell where the number of neighboring cells with density larger than $n_{shield}$ is greater than zero and less then 26 and the amount of heat it receives is proportional to $(q \Lambda_{0} N_{open})/26$, where $N_{open}$ is the number of cells in the surrounding of the considered cell which have a density lower than $n_{shield}$. 

In this paper, we shall analyze a number of models for which the combination of free parameters leads to TI, with $\beta=0.25$, $q=0.1$ and $n_{shield}=25\bar{n}$ where $\bar{n}$ is the mean particle density of the gas. This choice of parameters leads to TI where the thermal-equilibrium pressure decreases when density increases, and this for the whole temperature range (See Eq. 9 in V\'{a}zquez-Semadeni et al. 2000). However, with the imposed upper and lower temperature limits, gas can be forced to be in thermal equilibrium (i.e., cold gas at the minimum temperature $T_{min}$ is not allowed to cool below $T_{min}$ if its cooling rate is larger than its heating rate; the same arguments applies for warm gas which gets trapped at $T_{max}$. We vary the cooling rate $\Lambda_{0}$ in the different simulations. The cooling rate is inversely proportional to the cooling timescale $\tau_{cool}$ (Dib et al. 2004) which can be parameterized as a fraction $\eta$ of the dynamical time : $\tau_{cool}= \eta~\tau_{dyn}$. Dib et al. (2004) have run a large number of simulations similar to those presented in this paper, but for decaying turbulence. In those simulations the mass fractions of cold ($T < 300$ K) and intermediate ($300 < T < 6000$ K) are checked after they have reached equilibrium (or pseudo-equilibrium) values. They found that TI always takes place for values of $\eta \lesssim 1$ (See. Fig 1. in their paper). For $\eta$ $\gtrsim 1$ (corresponding to small values of $\Lambda_{0}$) the kinetic energy is dissipated before TI comes into play and no significant over-densities can be produced. On the other hand, for very small values of $\eta$ (large $\Lambda_{0}$) the available reservoir of thermal energy is radiated extremely fast and the medium achieves everywhere its minimum temperature. The simulation then resembles an isothermal simulation with a $\it{rms}$ Mach number close to unity. S\'anchez-Salcedo et al. (2002) discussed the development of TI in 1D numerical simulations as a function of the parameter $\eta$ in the presence of forced flows. They show that TI always takes place when $\eta \lesssim 1$ regardless of the forcing strength. In our simulations involving cooling, the driving occurs on large scales and $\eta$ is smaller than one, thus enabling the development of TI. 

The turbulent medium initially fulfills everywhere the Jeans stability criterion. After cooling is turned on, the gas cools preferentially inside the high-density fluctuations which become even more dense due to the large induced pressure gradients (particularly in the case of the smaller values of $\eta$). Thus, perturbations evolve in a highly dynamical way. The first significant over-densities (a factor of a few 100) are produced by means of TI. Condensations disappear either under the effect of localized turbulence, merge to form denser and larger condenations, or eventually the effect of background heating which affects mostly their outer boundaries. In the case of a weak cooling (large value of $\eta$), the condensations that form do not reach the stage of gravitational collapse, but they continue to cool and condense in order to achieve a pseudo-pressure equilibrium with the warmer phase. In the models where TI is efficient (small value of $\eta$), some of the most massive clumps become gravitationally bound and begin to contract under the effect of gravity, leading to even denser structures with central densities of a few 10$^{3}$cm$^{-3}$. In a later stage, condensations start merging and form larger condensations which ultimately will end up into a unique dense object. In order to understand the separate effects of TI and GI, we have performed a simulation with $\eta=0.3$ but without self-gravity (e.g., model CO03T2). This simulation evolves in a similar way to model CO03G2 until the point where condensations with central densities of a few 100 cm$^{-3}$ are formed, but these condensations do not merge together and the medium is segregated into two phases coexisting under rough pressure equilibrium conditions. The time evolution of the PDFs in simulations CO03T2 and CO03G2, displayed in Fig.~\ref{fig2} and Fig.~\ref{fig3}, respectively, illustrates the difference between the two models and the role played by gravity. Whereas the medium in model CO03T2 settles in a pseudo-stationary bimodal distribution, the PDF of model CO03G2 shows an increasingly moving tail towards higher densities as dense strcutures accrete additional matter under the effect of gravity. The implications on how these different behaviors could be interpreted from \ion{H}{i} observations will be detailed in Section 6. Fig.~\ref{fig4} and Fig.~\ref{fig5} show snapshots of the density structure at $t=\tau_{dyn}$ after cooling is turned on for models with a strong (CO03G2) and weak (CO09G2) cooling, respectively. Whereas the highest density at $t=\tau_{dyn}$ in model CO09G2 is $\sim 50$ cm$^{-3}$, in model CO03G2, the peak density has already reached a value of $\sim 5\times 10^{3}$ cm$^{-3}$.  

\section{FROM PHYSICAL DATA TO OBSERVATIONS}

The detailed structure in the three dimensional space can not be compared directly to observations. Rather, observations of large scale structures of the ISM are presented as emission maps in different atomic or molecular lines. The relatively low average density of our simulated medium (0.5 cm$^{-3}$) will not show up in emission maps of the high density tracers such as CO, CS and N$_{2}$H$^{+}$ and despite the fact that we have some detectable emission in the CO(1-0) line, the number of cells where this signal can be registered is very small. To make a direct comparison with observations possible, we calculate emission maps in the hyperfine structure line of the neutral hydrogen. This transition occurs for relatively low densities in the range of 0.1-1 cm$^{-3}$. We assume the number density fraction of hydrogen atoms to be $X=0.90$. Following a similar procedure to the one used in G\'{o}mez \& D'Alessio (2000) and Ballesteros-Paredes \& Mac Low (2002) but for the 21 cm \ion{H}{i} line instead of molecular emission lines, emission maps are built by solving the radiation transfer equation along each line of sight under the assumption of local thermodynamical equilibrium (LTE), 

\begin{equation}
\label{eq1}
\frac {dI_{\nu}} {ds}  = \kappa_{\nu} (S_{\nu}-I_{\nu}), 
\end{equation}

where $I_{\nu}$ is the intensity at the considered wavelength, $S_{\nu}$ is the source function which we replace by a blackbody radiation function (i.e., Planck function), $S_{\nu}=B_{\nu}(T)$, under the assumption of LTE where the temperature $T$ is defined by the local temperature of each cell, and $ds$ is the differential distance element along the line of sight. $\kappa_{\nu}$ stands for the absorption coefficient of the line. $I_{\nu}$ is determined by :

\begin{equation}
\label{eq2}
I_{\nu}  = I_{0} e^{-\tau_{\nu,tot}}+ \int_0^{\tau_{\nu,tot}} B_{\nu} e^{-\tau_{\nu}}d\tau_{\nu}, 
\end{equation}

and 

\begin{equation}
\label{eq3}
\tau_{\nu}  = \int_s^{\infty}\kappa_{\nu} ds  
\end{equation}

 where $\tau_{\nu}$ is the value of the local optical depth for the considered frequency and $\tau_{\nu,tot}$ is the total optical depth integrated along the box length. On our discrete grid, the integration in Eq.~\ref{eq2} is replaced by a summation, running over all the cells along the line of sight. In the general case, the absorption coefficient, $\kappa_{\nu}$ can be described by 

\begin{equation}
\label{eq4}
\kappa_{\nu}  = \frac {A_{ij} c^{2}}{8 \pi \nu^{2}} \frac{g_i}{g_j}\left[1-exp\left(-\frac{h \nu}{k_{B}T}\right)\right] n_j \phi(v), 
\end{equation}

where $A_{ij}$ is the Einstein transition coefficient when the local density exceeds the threshold density for excitation and which is zero if the density is less then this threshold. We choose a value for this threshold density\footnote{The value of the critical density in Rohlfs \& Wilson (1996) is 1 cm$^{-3}$, but observations in the \ion{H}{i}  21 cm line suggest that this value could be smaller (Muller et al. 2003), as they could infer densities of 0.06 cm$^{-3}$. Thus, the value of 0.5 cm$^{-3}$ should be regarded as a characteristic value.} of 0.5 cm$^{-3}$. $n_j$ is the number density of the atoms in the upper level $j$. $\phi(v)$ is the line profile, which is assumed to be thermal. In the particular case of the 21 cm \ion{H}{i} line, Eq.~\ref{eq4} can be simplified by replacing the exponential term by the first two terms of the Taylor expansion (Rohlfs \& Wilson 1996) such that

\begin{equation}
\label{eq5}
\kappa_{\nu}  = \frac {3 c^{2}} {32 \pi} \frac{1}{\nu_{10}} A_{10} n_{H}\frac{h} {k_{B} T} \phi(v), 
\end{equation} 

 where $n_{H}$ is the neutral hydrogen number density and is given by $n_{H}=X n_{tot}$ where $n_{tot}$ is the total number density of the gas. The Einstein transition coefficient and the frequency of the 21 cm \ion{H}{i} line are given by $A_{10}$=2.86888 $\times$ 10$^{-15}$ s$^{-1}$ and $\nu_{10}$=1.420405751786 $\times$ 10$^{9}$ Hz, respectively. $\phi(v)$ is given by 

\begin{equation}
\label{eq6}
\phi(v)  = \frac {c} {\nu_{10}} \left( \frac {m_H}{2 \pi k_B T}\right)^{1/2} exp\left[-\frac{m_H v^2}{2 k_B T}\right]
\end{equation}

where $m_H$ is the mass of the hydrogen atom. We tested the approach presented in Eqs.~\ref{eq1}-\ref{eq6} to analytical solutions that can be be obtained when simplifying assumptions are made such as assuming a uniform temperature and uniform density. In the latter case, the analytical solution is $I_{\nu}=I_{0} e^{-\tau_{\nu,tot}} + B_{\nu}(T) (1-e^{-\tau_{\nu,tot}})$. We checked for the particular case where $I_{0}=B_{\nu}(T)$ which leads to the solution $I_{\nu}=B_{\nu}(T)$. The two dimensional intensity obtained after solving for Eq.~\ref{eq2} is the intensity at the surface of the data cube. In order to mimick observational 21 cm \ion{H}{i} line maps, we still need to apply two corrections. The first one consists in convolving the result obtained by means of Eq.~\ref{eq2} with a function $G$ which represents the radio telescope response. For simplicity, we mimick observations with a single dish radio telescope and we describe $G$ by a two dimensional Gaussian function as displayed in Eq.~\ref{eq7} and Eq.~\ref{eq8}.

\begin{equation}
\label{eq7}
I_{\nu,o} = I_{\nu} \otimes G
\end{equation}

and 

\begin{equation}
\label{eq8}
G = \frac {1}{2\pi\sigma^2} exp\left[ -\frac{N_{x}^2+N_{y}^2}{2\sigma^2} \right],
\end{equation} 

The standard deviation $\sigma$ of $G$ is linked to the beam size $BS$ through the relation $BS = 2\sqrt{2\times ln2} \sigma$. The beam size is in turn fixed by the radio telescope's resolution $R$ and the angle $\Phi$ under which each cell is viewed such that $BS=R/\Phi$. $\Phi$ is given by the relation $\Phi=L_{b}/(N D)$ where $L_{b}$ is the physical size of the simulation box and $N$ the number of cells in one dimension. $D$ is the distance to the box (i.e., to Ho II). $D$ is fixed at a value of 3.38 Mpc (Karachentsev et al. 2002). We also set $R$ such that it represents a resolution of 4.5$\arcsec$/pixel which is the resolution of the high resolution \ion{H}{i} map of Puche et al. (1992) to which we intend to compare our results. $N_{x}$ and $N_{y}$ are defined as the number of grid cells present in the beam in each direction. They are obtained by dividing the beam size $BS$ by the angular size of one cell and in our case they are equal. The number of cells we have in the beam in each direction is $N_{x}=N_{y}=9$ which corresponds to a smoothing length value of 9$\times$7.8=70.2 pc. The smoothing length in the high resolution map of Puche et al. (1992) is $\sim$ 73 pc. Hence the beam size values in our simulations and in the observational map are quite similar. The second effect is the geometrical dilution of the intensity. The intensity at the distance $D$ is equal to the intensity at the surface, multiplied by a dilution factor $w=\Omega/2\pi\approx L_{b}^{2}/(2 N^{2} D{^2})$, where $\Omega$ is the solid angle which sustains one cell. One should also mention that because of the convolution, the smoothing of pixels which lie at the edges of the intensity map can be incorrect as they have no counterparts in all directions. We solve this problem by duplicating our map nine times and aligning the copies such that they form a $3 N\times3 N$ map on which we perform the convolution. Afterwards, we truncate $N\times N$ pixels inside the actual map to recover a smoothed complete $N\times N$ pixels map.

\section{THE OBSERVABLE QUANTITY}

The algorithm described in \S 3 has been applied to our simulations to compute the corresponding 21 cm \ion{H}{i} intensity maps. Examples of maps from two snapshots in two different models (i.e., Fig.~\ref{fig4} and Fig.~\ref{fig5}) are displayed in Fig.~\ref{fig6} and Fig.~\ref{fig7}, respectively. The value of the intensity is given in Jansky and a background intensity $I_{0}$ of $10^{4}$ Jansky is assumed. The maps at $t=\tau_{dyn}$ show a structured distribution in the 21 cm line in the form of shells and holes similar to the observations (compare to Fig. 1 in Rhode et al. 1999 and Fig. 1 in Kim et al. 1999). The stronger pressure gradients at the interfaces of the dense clouds in models with more efficient cooling (smaller values of $\eta$) lead to more compact filaments and shells in the three dimensional structure which is also seen in the 21 cm \ion{H}{i} maps. In order to quantify the comparison between the models and the observations, we calculate the autocorrelation function, which is defined as follow, 

\begin{equation}
\label{eq9}
 S_2(\Delta x,\Delta y) =  \frac {\sum_{i,j}^{L}(I_{\nu,o}(i,j)-\overline{I_{\nu,o}})(I_{\nu,o}(i+\Delta x,j+\Delta y)-\overline{I_{\nu,o}})} {\sum_{i,j}^{L} (I_{\nu,o}(i,j)-\overline{I_{\nu,o}})^2}
\end{equation}

where $\overline{I_{\nu,o}}$ is the average value of the intensity and $\Delta x$ and $\Delta y$ represents the distance of cells in both directions. There are different ways to deal with the cases when the distance  $\Delta L=\sqrt{{\Delta x}^2 + {\Delta y}^2}$ extends beyond the box (or sub-box) edges in either directions. One could assume, at every scale, a self-periodicity of the \ion{H}{i} in every directions. However this is not physically justified. The ideal treatment would be to use for each box (or sub-box) its natural neighbors for the calculation of the autocorrelation length. Such an approach would be favorable to small scales where natural boundary pixels are available but would still imply a self-periodicity for the large scales. We took the option of treating all scales equally, by using in the calculation of $S_{2}$ only those pixels which reside inside the box at the given scale. The price of such an approach, though it yields more consistent results, is to increase the statistical error bars since a smaller number of pixels is involved in the calculation of $S_{2}$ at the larger scales. $S_{2}$ has a maximum of $S_{2}=1$ for $\Delta L=0$ and drops for larger scales. The characteristic value of the autocorrelation length $L_{cr}$ can be reasonably defined as being the physical scale for which the autocorrelation function has dropped substantially below $1$. We choose $L_{cr}$ to be the value of the length scale when $S_{2}$=0.2. The autocorrelation length is then defined as the average of the three autocorrelation length values derived from \ion{H}{i} maps calculated for the three projections of the simulation box. Houlahan \& Scalo (1990) pointed out that the estimate of $S_{2}$ on small scales might be affected by structures with characteristic scales that are larger than the box size. this indeed might happen for some of the smaller boxes, if a box with the smallest size we consider (125 pc) is centered on a condensation which is larger than the box itself. In that case the estimate of $L_{cr}$ might be dominated by the self terms and would not be accurate. However, on those small scales (125 pc), we calculate, at each time step, the value of $L_{cr}$ $8\times8\times3=92$ times on independent sub-boxes and many of those sub-boxes are not centered on dense condensations which happen to be larger than them. So, due to the large statistics, we are confident that we pick up the signature of cross terms (undoubtedly always affected by the self terms) even on the smallest scales we consider. In order to confirm this statement, we have checked, at the smallest scale we consider (i.e., $L=125$ pc $\sim 16$ cells), what is the fraction of points that have $I_{\nu}(i-8,j) < I_{\nu}(i,j) < I_{\nu}(i+8,j)$ and $I_{\nu}(i-8) > I_{\nu}(i,j) > I_{\nu}(i+8))$, where $I_{\nu}$ is the emissivity and i and j the indices in each direction. Similarly, we have also checked the fraction of points that fulfill the same criterion in the $j$ direction. By doing so, we can evaluate what is the fraction of points that are part of a gradient on that scale. The closer to unity the fraction of points which fall in this category is, the lesser the probability that independant structure or substructure exists on that scale. We find, for directions $i$ and $j$, respectively, the following fractions at $t=\tau_{dyn}$ : for $\eta=0.3$ 0.49 and 0.44, for $\eta=0.6$ 0.54 and 0.55 and for $\eta=0.9$ the values of 0.64 and 0.58. The fractions are closer to one for the weaker efficiency cooling models which is seen in the form of smoother structures in Fig.~\ref{fig7} as compared to Fig.~\ref{fig6}. These numbers are clearly smaller than one and indicate the existence of density fluctuations on scales smaller than 125 pc . 

\section{THE PHYSICAL SIZE-CORRELATION LENGTH RELATION IN Ho II}      

In order to check the validity of the comparison of $L_{cr}$ on a particular scale between the observations and the synthetic maps, the dependence of $L_{cr}$ on the size $L$ of the analyzed region must be checked in Ho II. For a given scale $L$, we subdivide the observed \ion{H}{i} map of Ho II in side by side boxes, each of size $L$, excluding pixels within a distance $L$ from the edges of the map to avoid contamination by the outer pixels which might not belong to Ho II. For larger scales we additionally consider squared regions that are shifted by a third of the box size in order to increase the number of boxes on which we evaluate $L_{cr}$. The characteristic autocorrelation length at a given scale is calculated by averaging the individual values. The error is simply assumed to be the standard deviation with respect to the average value at the considered scale. The $L-L_{cr}$ relation we find for Ho II is displayed in Fig~.\ref{fig8}. It shows a linear growth of $L_{cr}$ with $L$ up to a scale of $\sim 6$ kpc and, for larger scales, a nearly constant value, independent of the physical scale. The shape of the $L-L_{cr}$ relation suggests that the structure is correlated for scales less than $\sim 6$ kpc. On larger scales, structures in different regions seem to form independently. The relationship between the position of the turnover point in the $L-L_{cr}$ relation and the length scale of the driving mechanism as well as the value of the slope in the linear regime is what we attempt to explore in the next section using comparisons between the observed and simulated \ion{H}{i} maps of Ho II.  For the $L-L_{cr}$ relation in Ho II, a linear fit to the data points for scales ranging between 0.25 kpc and 6 kpc yields a slope of $0.284\pm0.061$ (i.e., fit over-plotted to the data in Fig.~\ref{fig6}).

\section{COMPARING THE MODELS TO THE OBSERVATIONS} 

\subsection{COMPARISON TO THE ISOTHERMAL MODELS}
 
We first analyze the set of isothermal models in which the turbulence is driven at different length scales and at different Mach numbers. These are labeled  ISOimj in Table. 1,  where i stands for the wavenumber on which the energy is injected and j the Mach number. The purpose of these simulations is to check if the observed \ion{H}{i} hole and shell like structure can emerge as a result of turbulence alone. Fig.~\ref{fig9} shows the $L-L_{cr}$ relations derived for models ISO2m1 and ISO4m1. The over-densities $\delta\rho/\rho$ created locally by converging flows do not exceed the value of 3 in these simulations which are characterized by a Mach number $Ma=1$. Fig.~\ref{fig9} shows the $L-L_{cr}$ relations derived for models ISO2m1 and ISO4m1. The values of $L_{cr}$ are obtained by averaging over time for time dumps between $\tau_{dyn}$ and $2\times \tau_{dyn}$ (after a dynamical equilibrium of the system is reached; i.e., similar PDFs, see Fig.~\ref{fig1}) and the three dimensions of the simulation box. In these simulations the turbulence is driven with a wave number of $k=2$ and $4$, respectively (i.e., $L_{driv}$ of 0.5 kpc and 0.25 kpc, respectively). In both models the $L-L_{cr}$ relation shows a linear increase of $L_{cr}$ with $L$ up to a scale which is equal to $L_{driv}$. For scales larger then $L_{driv}$ the $L-L_{cr}$ relation nearly flattens. The turnover point in the $L-L_{cr}$ relation seems to be associated with the scale of the driving mechanism. Therefore, it can only show up if $L_{cr}$ is calculated from small scales up to scales that are larger than $L_{driv}$. A comparison of the models and the observations at a given scale is meaningful, only if that scale lies below the turning point of the $L-L_{cr}$ relation both in the models and the observations. A small increase in $L_{cr}$ is still visible with increasing scale. This is probably due to the presence of a non zero upward energy cascade which distributes some of the injected energy on scales larger than $L_{driv}$. The direct comparison of Fig.~\ref{fig8} to Fig.~\ref{fig9} implies that energy is injected into the ISM of Ho II on a scale of the order of $\sim$ 6 kpc. This clearly favors a large-scale driving rather than a driving by supernova explosions. The kinetic energy power spectrum drawn by Stanimirovi\'c \& Lazarian (2001) for the Small Magellanic Cloud (SMC) speaks for a large scale driver operating in the SMC as well. The spectrum shows no signs of energy injection up to the largest scale they have considered (4 kpc). The linear regime below the turning point is governed by the energy decay of the turbulence from large to small scales. Thus, the slope of the $L-L_{cr}$ relation in that regime is an indication of the gas equation of state. A linear fit to the $L-L_{cr}$ data for scales smaller or equal to $L_{driv}$ in models ISO2m1 and ISO4m1 yields a slope of $0.209\pm 0.052$ and $0.201\pm 0.096$, respectively. This is a factor $\sim 1.4$ smaller than the value yielded by the observations and is probably an indication that the hypothesis of isothermal gas is not entirely valid.

 We now check if these conclusions still hold when turbulence is driven at different $rms$ Mach numbers. We compare the observations to models ISO2m05, ISO2m1, ISO2m2, ISO2m3 and ISO2m5 where turbulence is driven with Mach numbers of 0.5, 1, 2, 3 and 5, respectively. In these simulations turbulence is driven with a wave number $k=2$ ($L_{driv}=0.5$ kpc) and thus, a comparison of these models to the observations is performed at the 0.5 kpc scale. Fig.~\ref{fig10} shows that the autocorrelation length remains independent of the strength with which turbulence is driven. 

\subsection{COMPARISON TO THE MODELS WITH COOLING AND HEATING}

We now turn to more realistic models where the cooling and the heating of the gas are present, as well as self-gravity. We should mention that the purpose of this section is not to recover the precise structure of the gas in Ho II, but rather to show qualitatively the importance of the combined effects of turbulence and the gas thermodynamics in generating the observed structures. The models have the ratios of the cooling time to the dynamical time of $\eta=0.3,0.6$ and $0.9$ (strong, intermediate and weak cooling). In these simulations turbulence is driven with $k=2$ ($L_{driv}=0.5$ kpc) and the comparisons to the observations are performed at the scale of 0.5 kpc. The advantage of comparing the autocorrelation length on the 0.5 kpc scale is that a large number of independent 0.5 kpc$^2$ squares can be selected from the \ion{H}{i} map of Ho II (we select $80\times 80$ 0.5 kpc$^2$ squares), thus enhancing the accuracy of the observed value at this scale. Though the observed value is only an estimate of the characteristic autocorrelation length at the time of the observations and although the holes and shells are continuously interacting, the selected sample is large enough to consider the observed value at the $0.5$ kpc scale $L_{cr,0.5} \sim 142$ pc as a realistic stable value. Due to the non-linear evolution of structures under cooling and gravity, a time average for the value of $L_{cr}$ is not meaningful, as it could smear out the true behavior of the gas. In order to distinguish the effects of TI and GI on the large scale structure as it would appear in the \ion{H}{i} map, we compare two models with the same cooling rate ($\eta=0.3$) but with and without self-gravity (models CO03G2 and CO03T2, respectively). Fig.~\ref{fig11} shows that the evolution of $L_{cr}/L_{driv}$ is nearly similar in both models until $t\sim1.7~\tau_{dyn}$, despite the fact that in the case of the run with gravity, the condensations that form following the early condensation process by TI are a few order of magnitude denser than in the case without self-gravity. Only when the larger and denser condensations start accreting the smaller ones, $L_{cr}/L_{driv}$ rises again. This indicates that the observed plateau in $L_{cr}/L_{driv}$ between $t\sim0.7~\tau_{dyn}$ and $t\sim1.7~\tau_{dyn}$ is a signature of TI alone. 

Fig.~\ref{fig12}.a compares the time evolution of the ratio $L_{cr}/L_{driv}$ in models with different values of $\eta$, with all these models having self-gravity included. In contrast to the above discussed high efficiency cooling case, the autocorrelation length derived from the model with $\eta=0.9$ (weak cooling, model CO09G2) is in good agreement with the observational value within the 1 $\sigma$ error bars, and this over a span of time which is $\sim 1.5~\tau_{dyn}\sim 150$ Myrs. This is comparable to the age of the oldest ($\sim~135$ Myrs) holes observed in Ho II when considering the sample of Rhode et al. (1999) of holes that do not host any remnant stellar cluster. Fig.~\ref{fig12}.b shows the $L-L_{cr}$ relation for model CO09G2 at $t=1.5~\tau_{dyn}$. The slope in the linear regime is $0.243\pm0.045$. In this model, the density in the densest clumps is only a few times larger than the average density. Here cooling is complementary to turbulence. When holes are created by turbulence, TI tends to enlarge those holes under the effect of the local pressure gradients at their boundaries. The curve corresponding to $\eta=0.6$ (model CO06G2) shows a good agreement with the observations for a time span of $0.8~\tau_{dyn}\sim 80$ Myrs. The system condenses further and $L_{cr}/L_{driv}$ drops to intermediate values between those of the high efficiency and week efficiency cooling until it reaches the stage of condensation-condensation accretion and the $L_{cr}/L_{driv}$ value increases again. Yet, the evolution of the \ion{H}{i} morphology if star formation occurs can not be followed with our models. However, the low star formation rate (SFR) of Ho II estimated to be in the range 0.00084-0.049 M$_{\odot}$ yr$^{-1}$ (Hunter \& Gallagher 1985; Stewart et al. 2002) suggests that many of the condensations that form due to TI are probably destroyed before they can form stars, possibly by the galactic shear\footnote{The galactic shear is however not very strong in Ho II. Using the velocity curve of Bureau \& Carignan (2002) we could estimate that, at the radii of 15 kpc, galactic shear acts on a timescale of $\sim 2\times10^{9}$ which is larger than the age of the observed \ion{H}{i} holes.} or other processes. This argues against strong cooling which tends to forms stars more efficiently and which imprints a smaller $L_{cr}$ value to the \ion{H}{i} morphology.

We have checked to which extent the results presented above depend on numerical resolution. For that purpose, we run two additional models similar to CO09G2 with the resolutions of $64^{3}$ and $192^{3}$. Fig.~\ref{fig13} displays the time evolution of $L_{cr}(500$ pc)$/L_{driv}$  where $L_{driv}=500$ pc in all models. Therefore the latter ratio corresponds to the value of the slope of the $L-L_{cr}$ in the 'linear' regime. The three models yield relatively similar values, in particluar models with the resolutions $128^{3}$ and $192^{3}$, whereas the model with the resolution $64^{3}$ shows larger fluctuations.
 
\section{CONCLUSIONS AND DISCUSSION}

\begin{figure}
\plotone{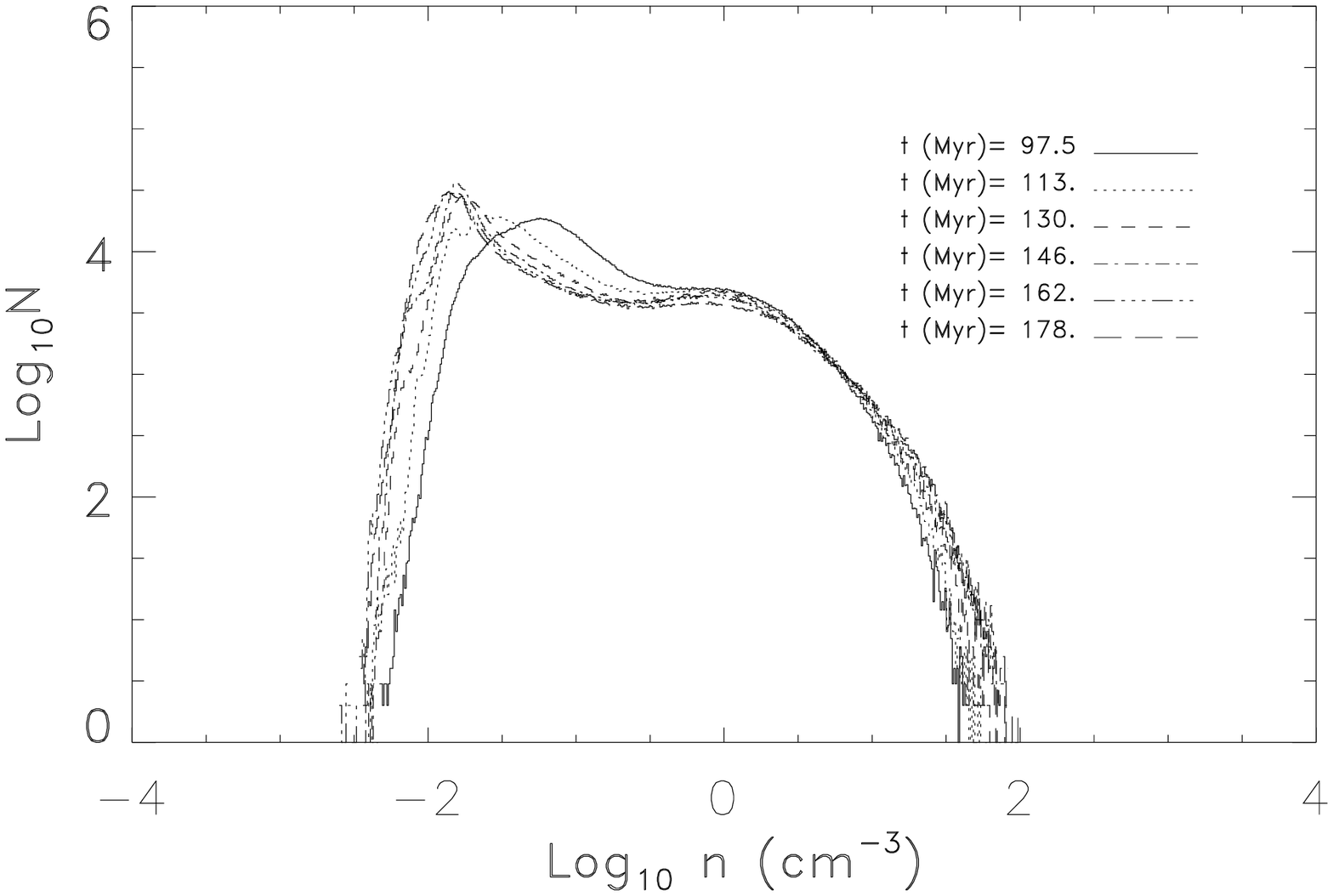}
\caption{Time evolution of the number density distribution function in simulation CO03T2.}
\label{fig2}
\end{figure}

\begin{figure}
\plotone{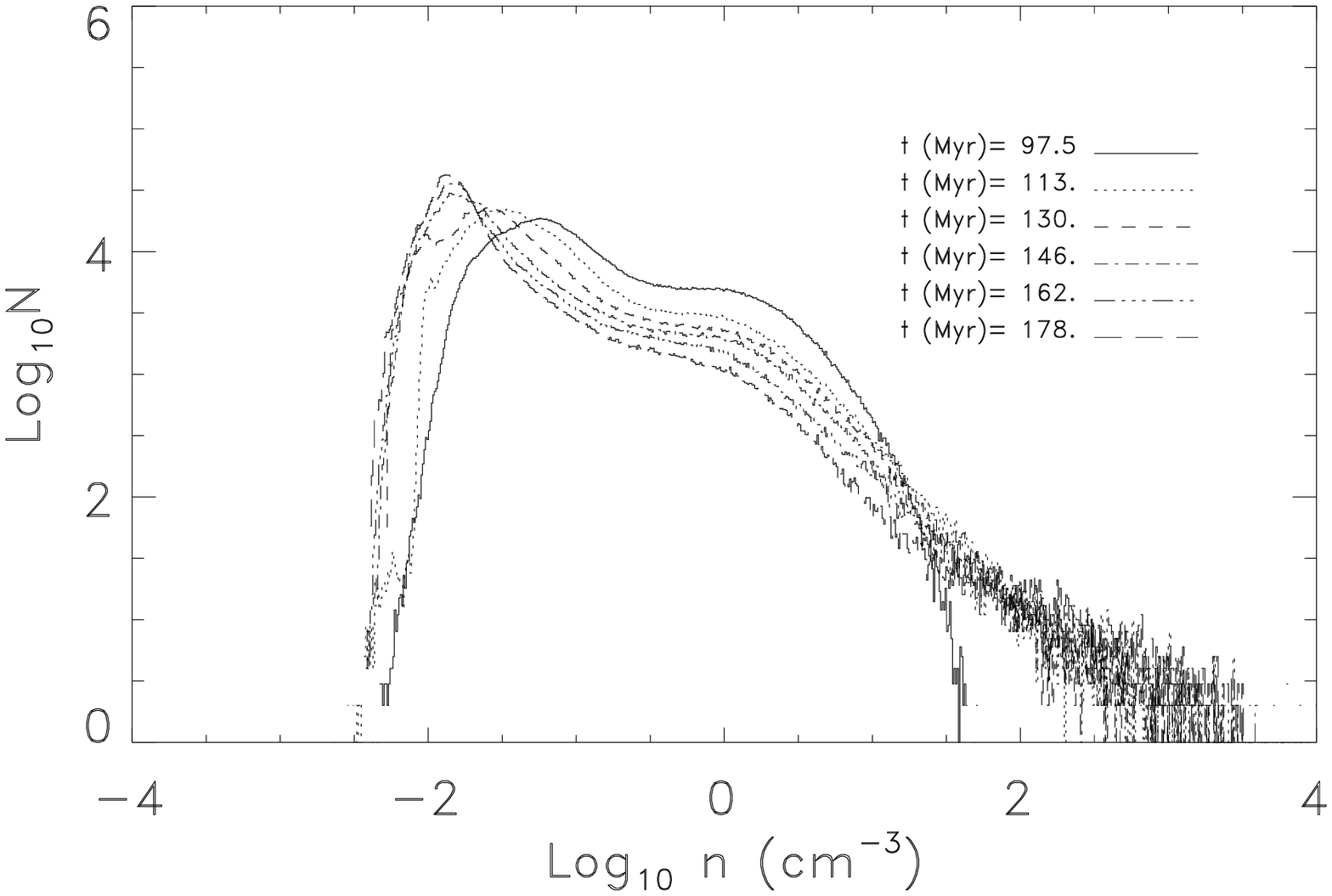}
\caption{Time evolution of the number density distribution function in simulation CO03G2.}
\label{fig3}
\end{figure}

\begin{figure}
\plotone{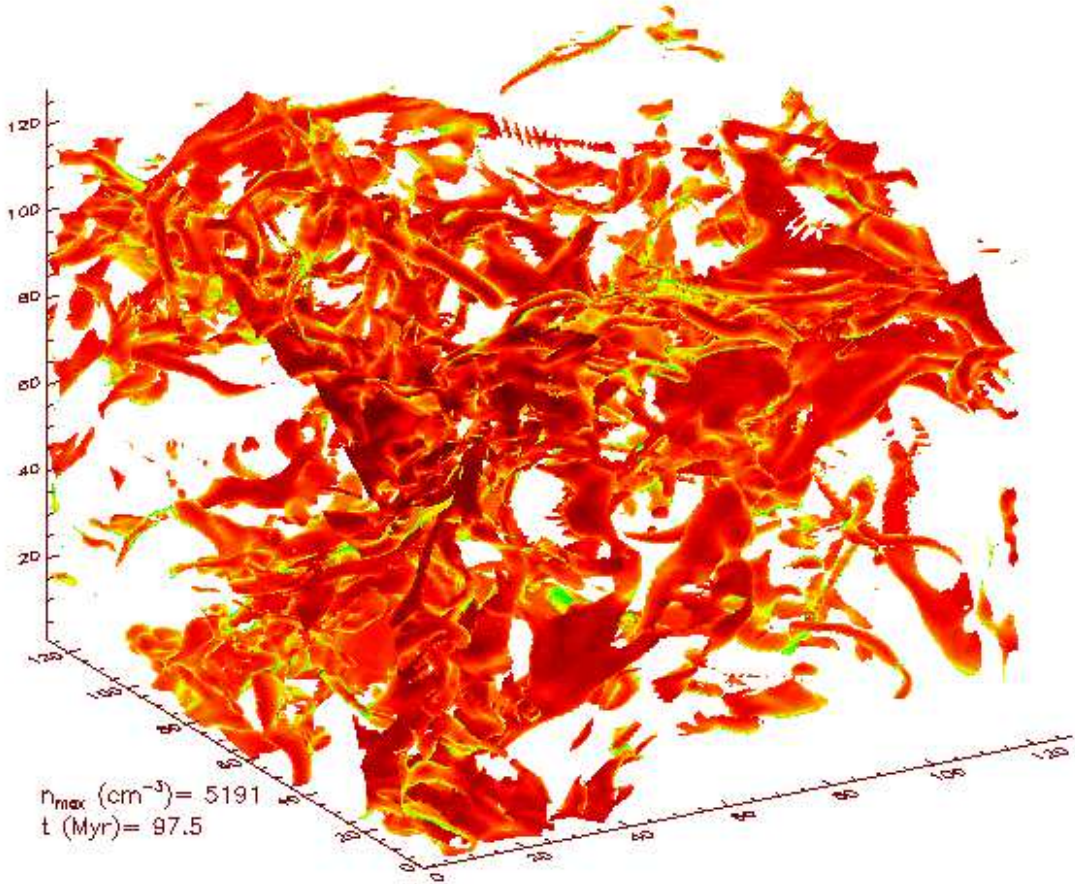} 
\caption{Three dimensional density structure of the medium at t=$\tau_{dyn}$ in model CO03G2. The iso-density surface corresponds to a value of 10 $\bar{n}$ ($\bar{n}=0.5$cm$^{-3}$). The axis labels correspond to the number of cells in each direction.}
\label{fig4}
\end{figure}

\begin{figure}
\plotone{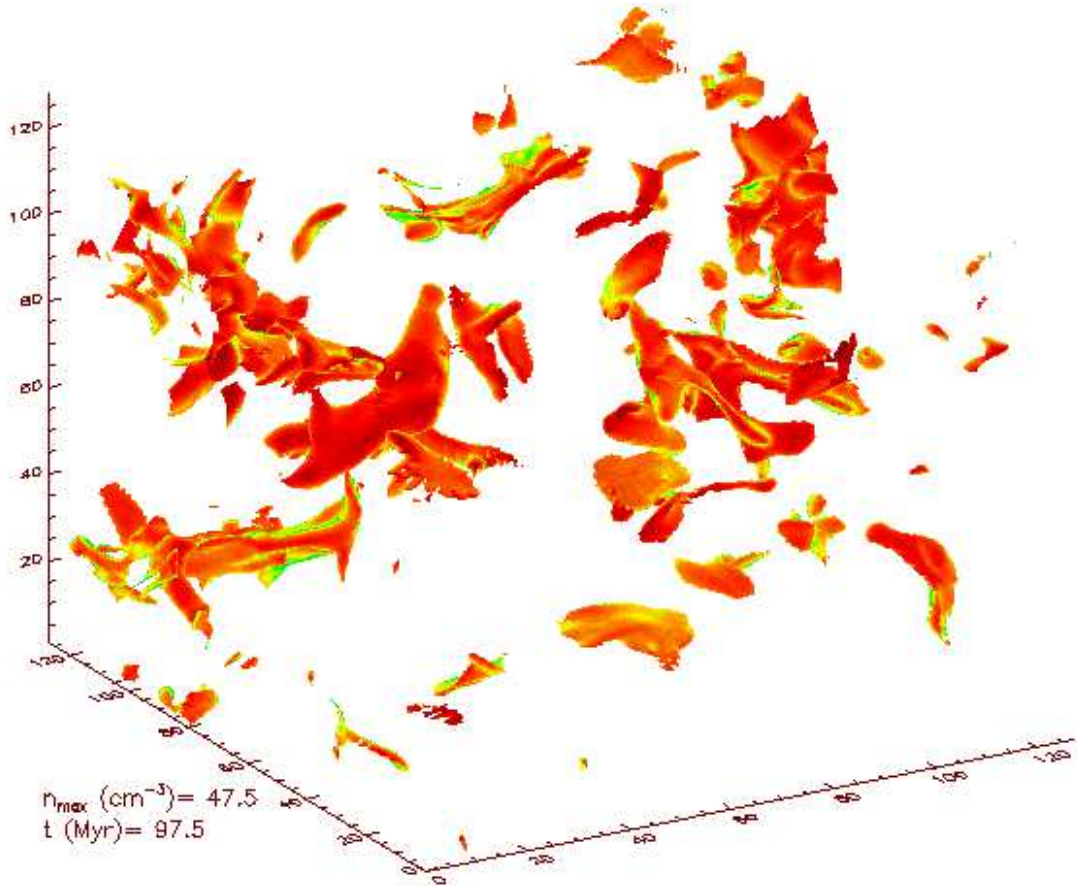}
\caption{Three dimensional density structure of the medium at t=$\tau_{dyn}$ in model CO09G2. The iso-density surface corresponds to a value of 2 $\bar{n}$. A smaller density threshold has been used in this model because of the lower over-densities reached in this model compared to model CO03G2.}
\label{fig5}
\end{figure}

\begin{figure}
\plotone{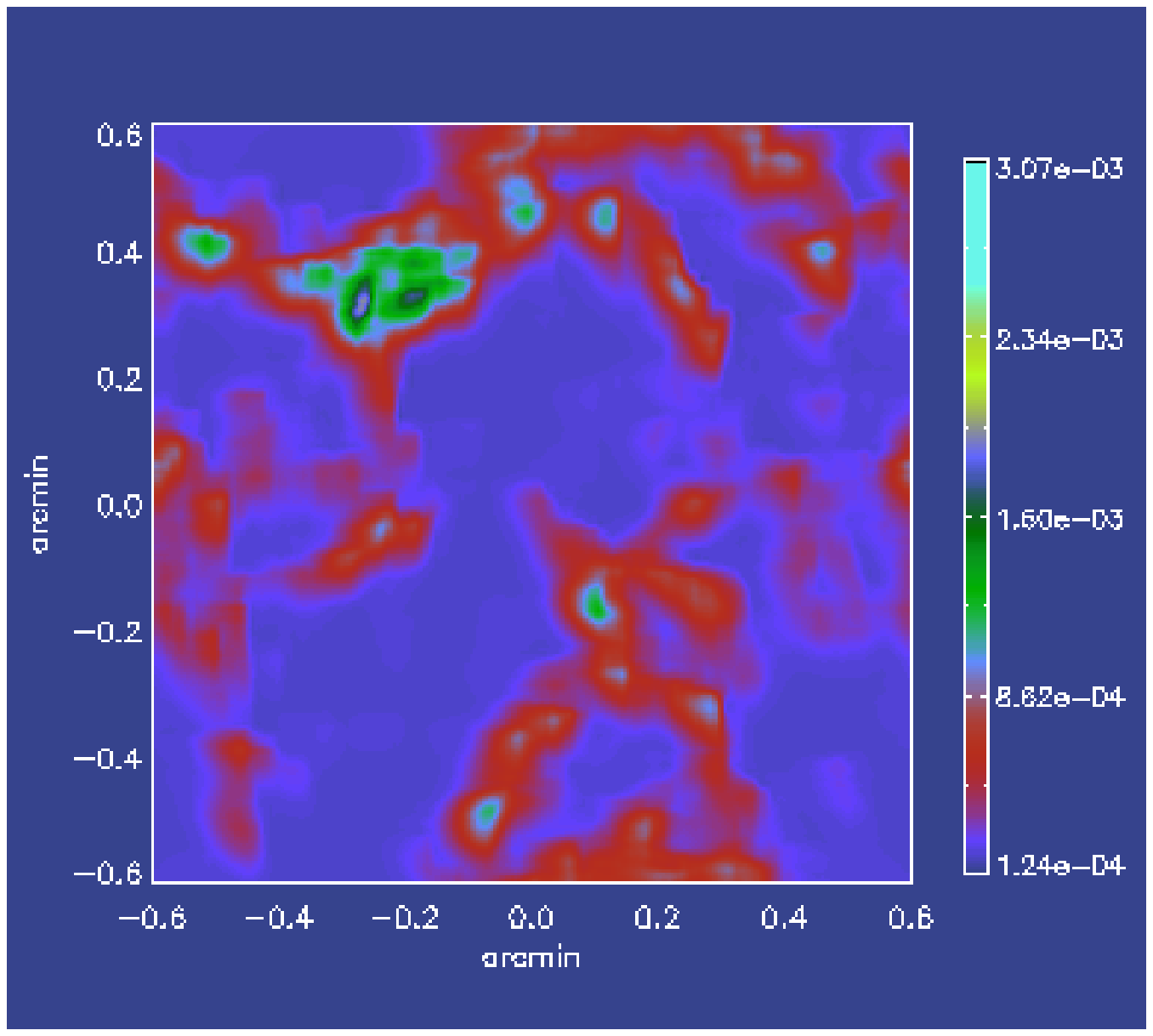}
\caption{21 cm \ion{H}{i} map of the medium at $t=\tau_{dyn}$ for model CO03G2. The value of the intensity is given in Jansky.}
\label{fig6}
\end{figure}

\begin{figure}
\plotone{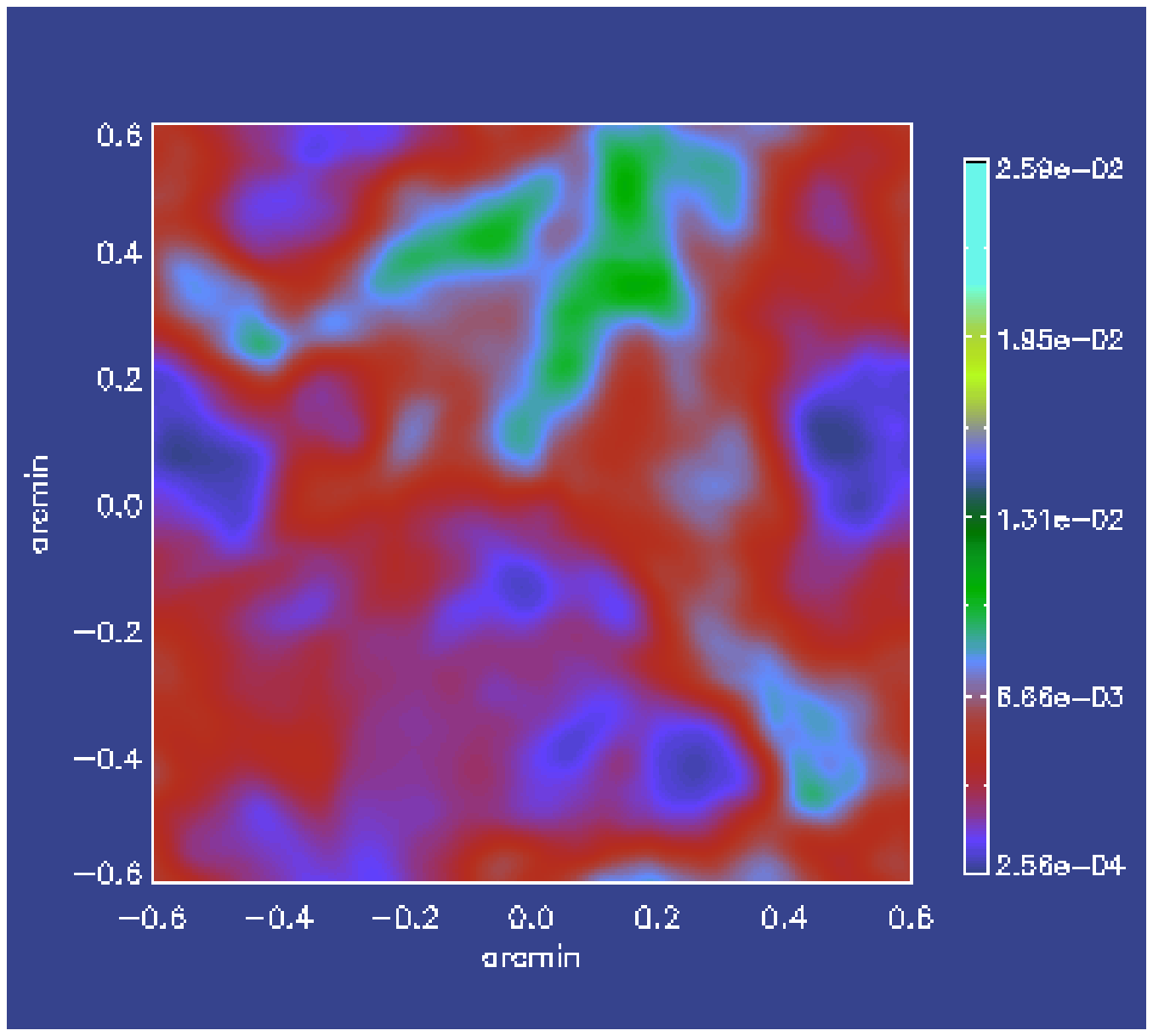}
\caption{21 cm \ion{H}{i} map of the medium at $t=\tau_{dyn}$ for model CO09G2. The value of the intensity is given in Jansky.}
\label{fig7}
\end{figure}

\begin{figure}
\plotone{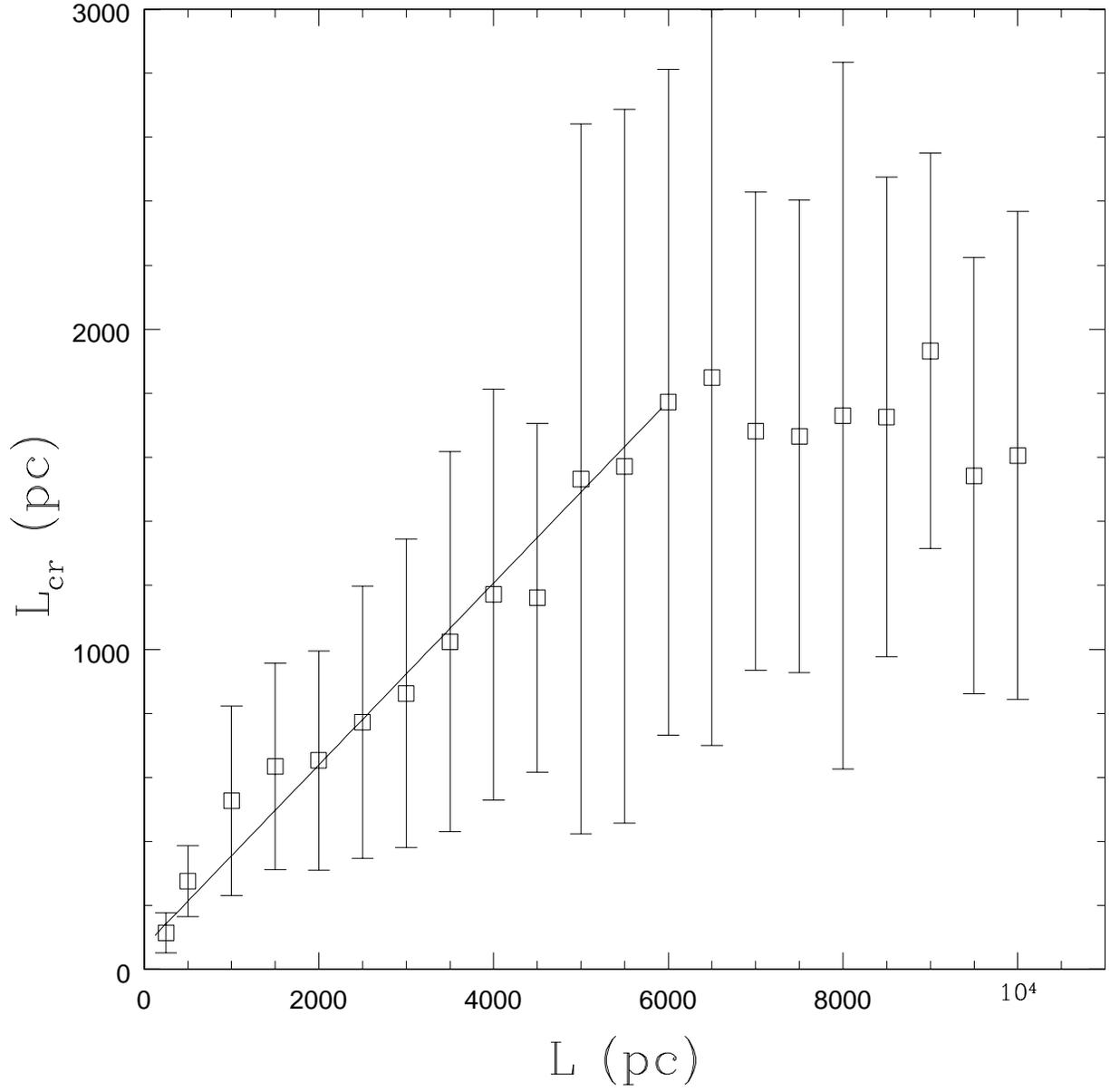}
\caption{$L-L_{cr}$ relation in Holmberg II.}
\label{fig8}
\end{figure}

\begin{figure}
\plotone{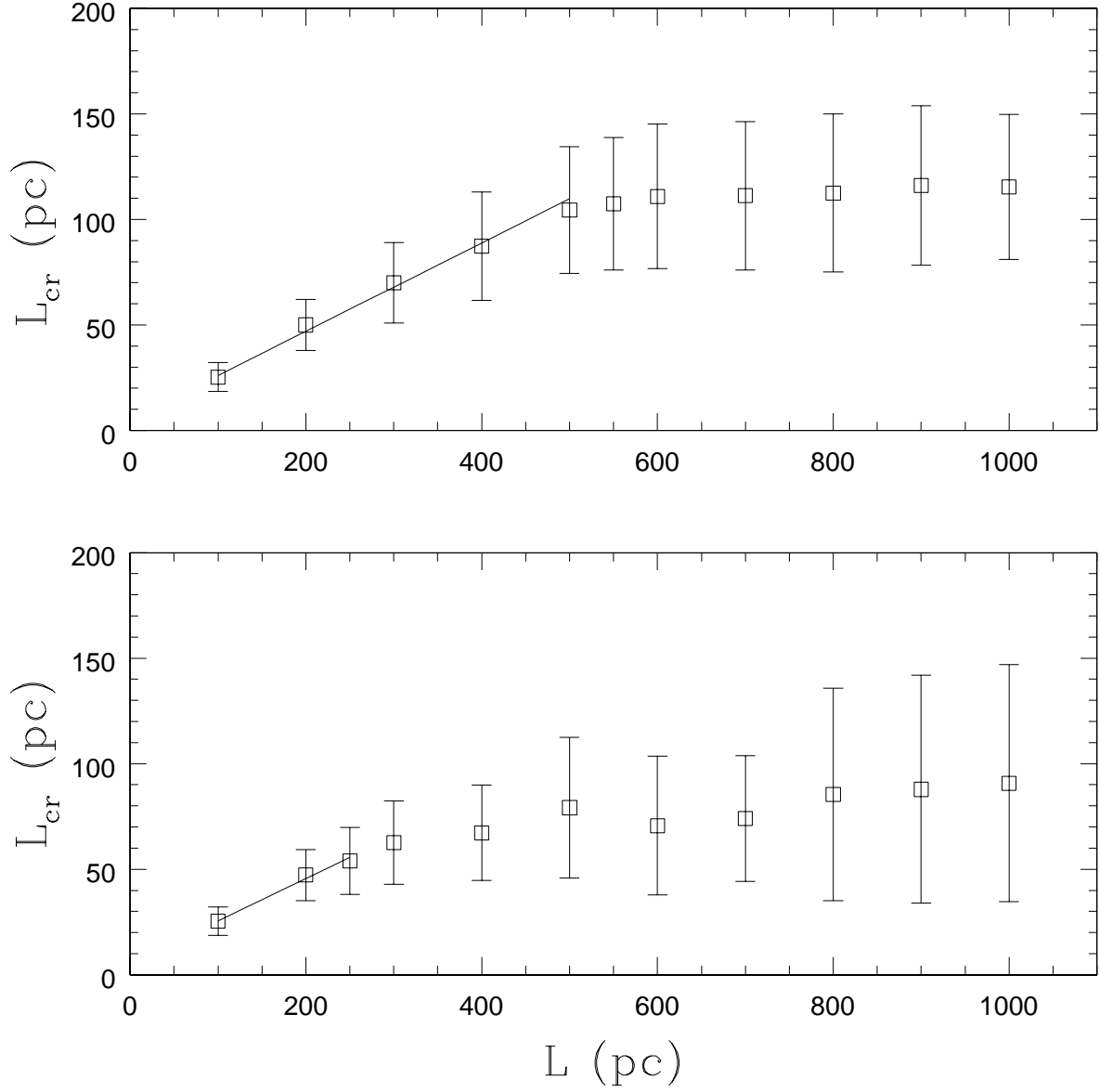}
\caption{$L-L_{cr}$ relation in models ISO2m1 (top) where the driving lengthscale is 0.5 kpc and ISO4m1 (bottom) with a driving lengthscale of 0.25 kpc.}
\label{fig9}
\end{figure}

\begin{figure}
\plotone{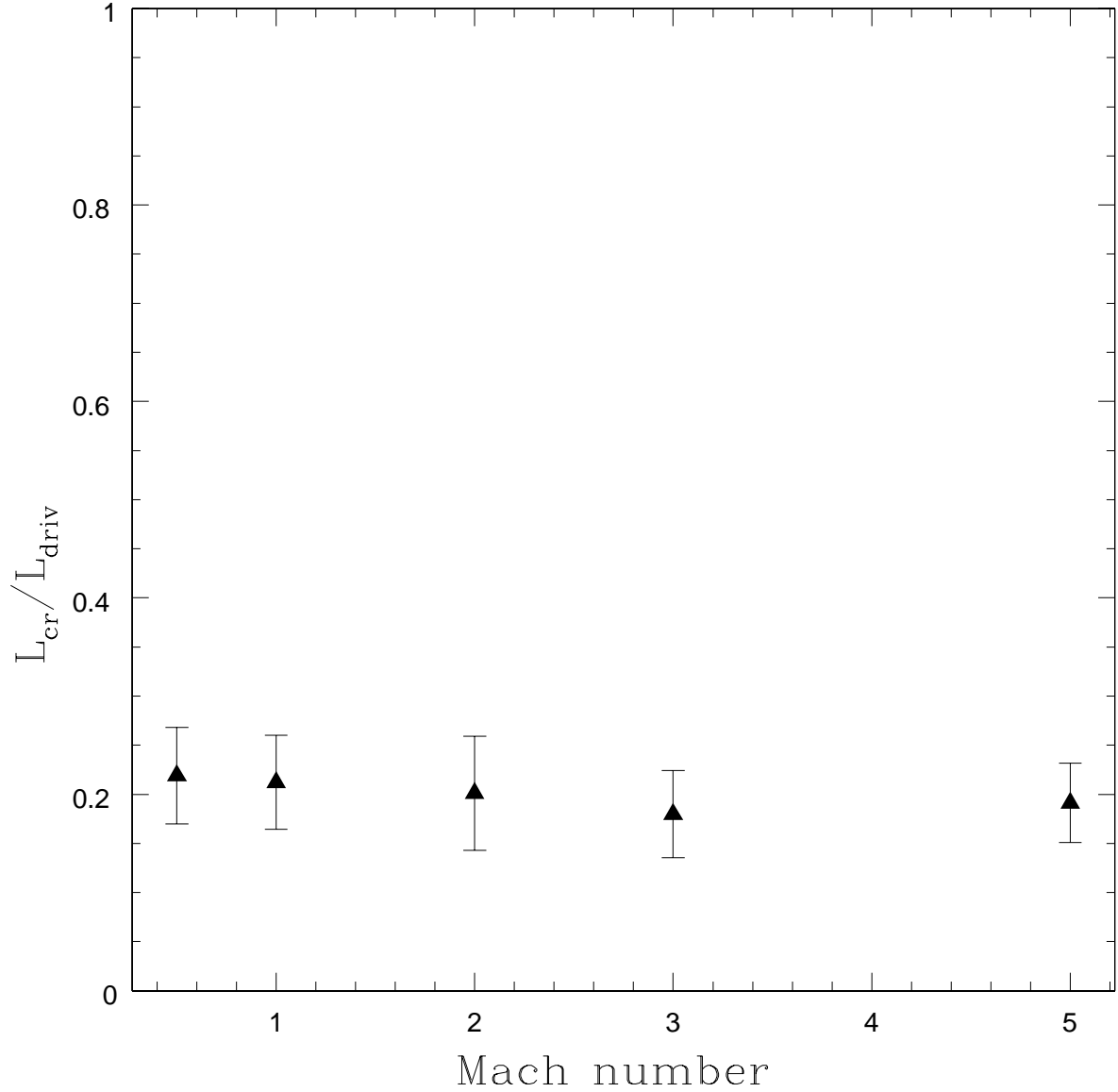}
\caption{Dependence of the autocorrelation length on the {\it rms} Mach number in the isothermal simulations. The driving wave number in these simulations is $k=2$ ($L_{driv}=0.5$ kpc).}
\label{fig10}
\end{figure}

\begin{figure}
\plotone{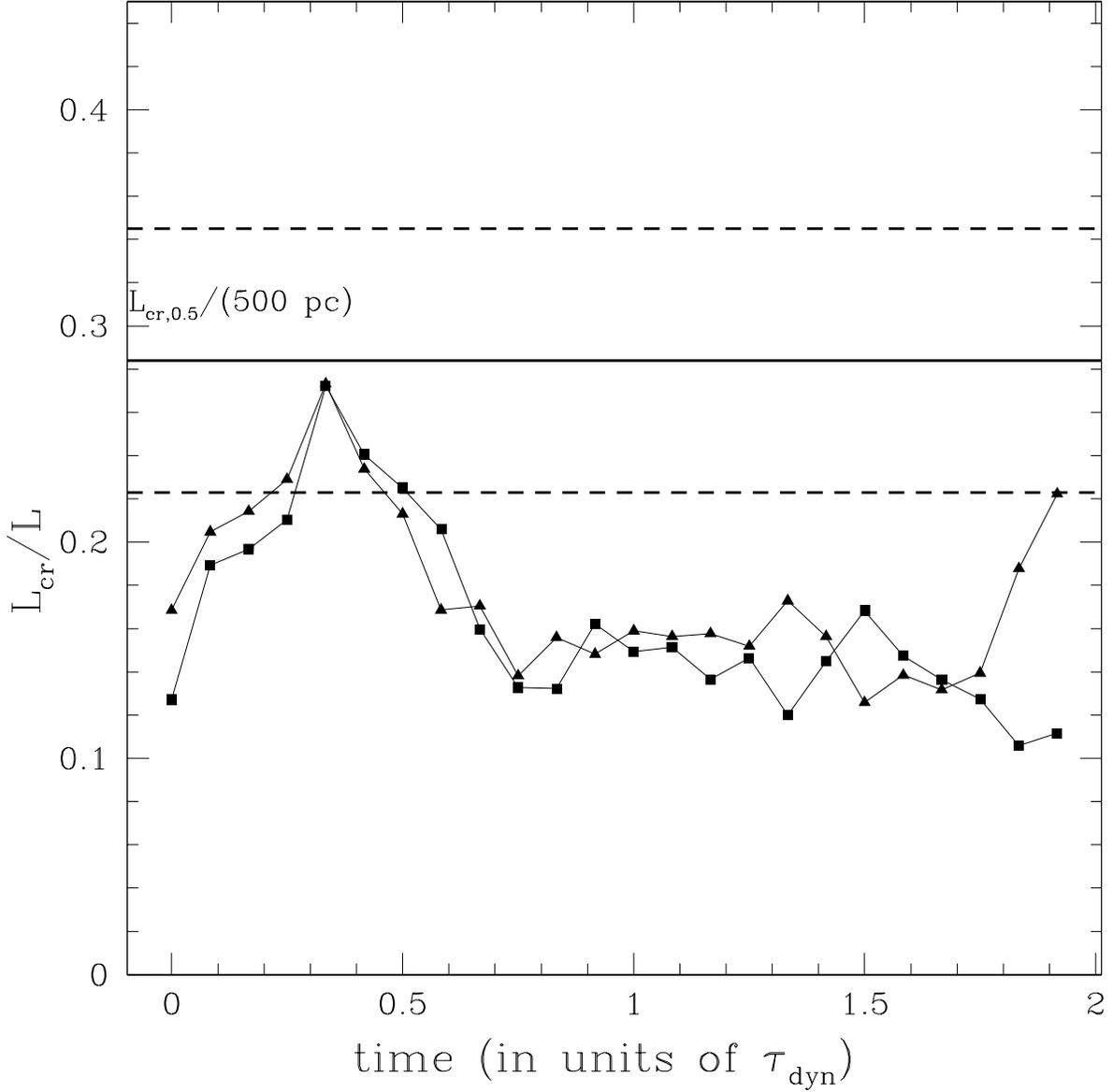}
\caption{A comparison of the time evolution of the autocorrelation length at the scale of 0.5 kpc between two models with self gravity (model CO03G2, full triangles) and without self-gravity (model CO03T2, full squares). $L_{cr}$ is normalized to the driving lengthscale in those models $L_{driv}=500$ pc, thus ($L_{cr}/500$ pc) corresponds to the slope of the $L-L_{cr}$ relation in the linear. non-constant, regime} 
\label{fig11}
\end{figure}

\begin{figure}
\plotone{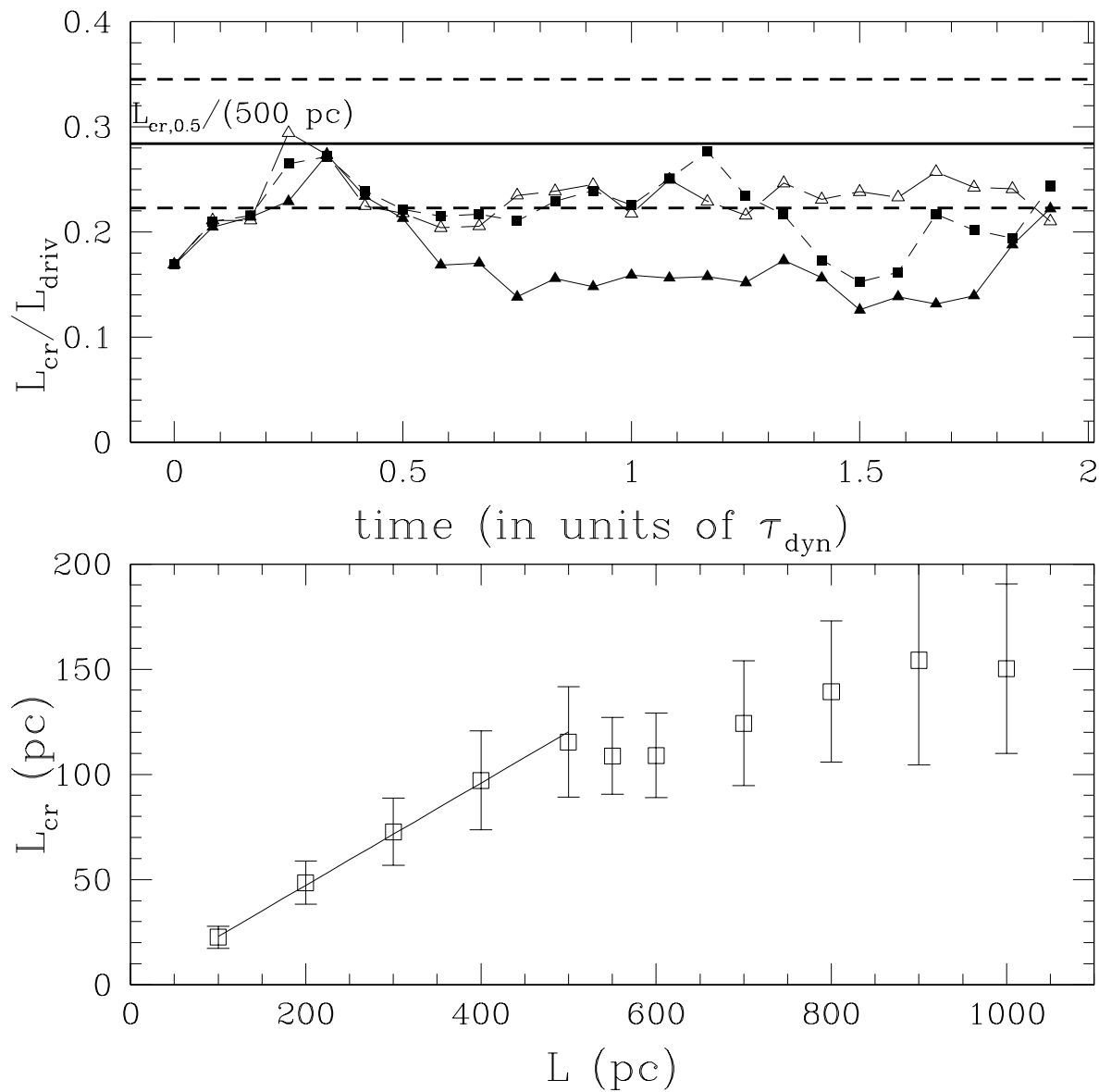}
\caption{a) (top) Time evolution of the autocorrelation length in models with cooling and gravity. the models are CO03G2 (full triangles), CO06G2  (full squares) and CO09G2 (open triangles). b) (bottom) $L-L_{cr}$ relation in model CO09G2 plotted at $t=1.5~\tau_{dyn}$.}
\label{fig12}
\end{figure}

\begin{figure}
\plotone{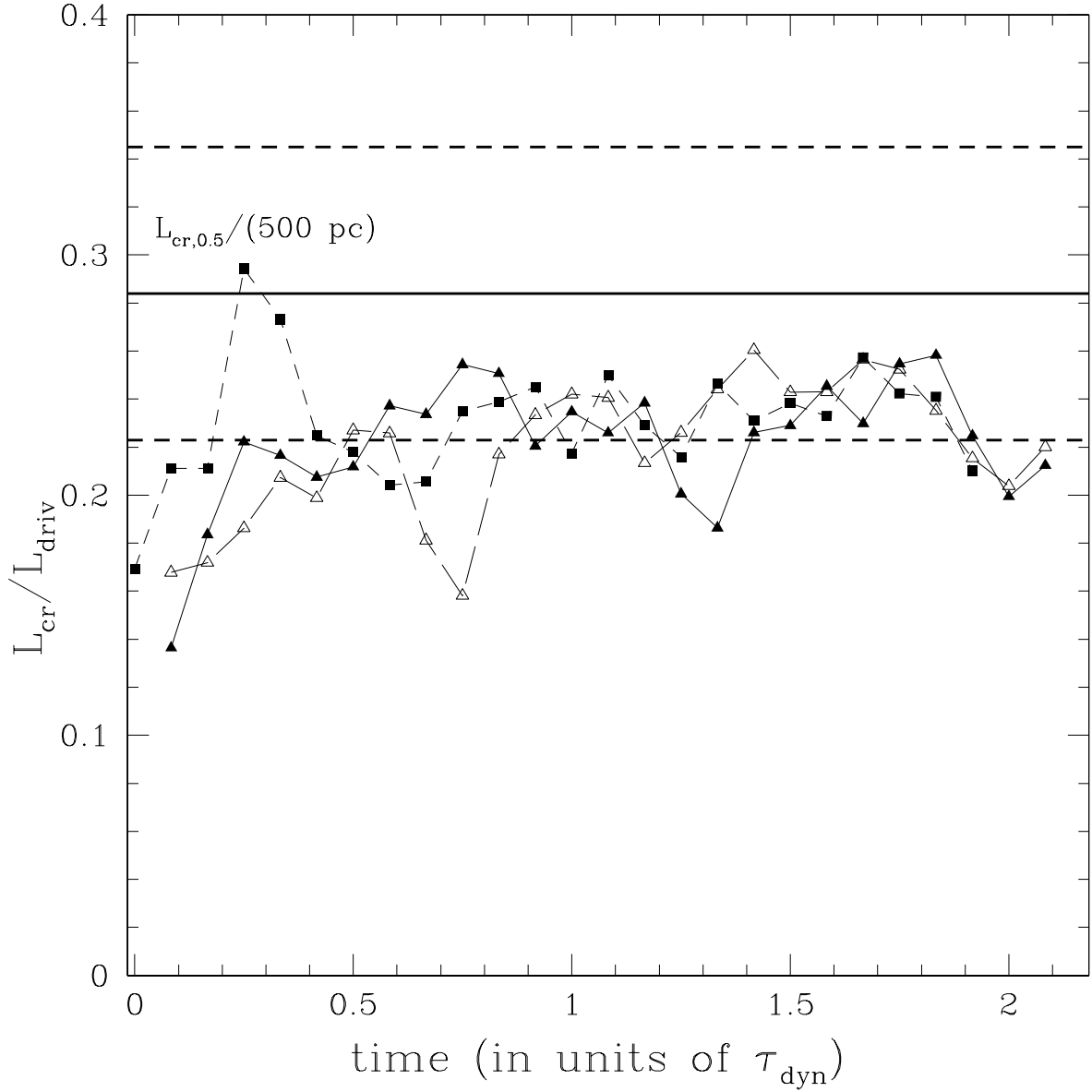}
\caption{Time evolution of the autocorrelation length in CO09G2 type models with resolutions of $64^{3}$ (full triangles), $128^{3}$ (filled squares) and $192^{3}$ (open triangle).}
\label{fig13}
\end{figure}

In this work we compare 21 cm \ion{H}{i} simulated maps to the high resolution \ion{H}{i} map of Puche et al. (1999) of the dwarf irregular, gas rich galaxy Holmberg II (Ho II). The maps are built by using a post processing of the radiative transfer equations on different models of 1 kpc scale local three dimensional hydrodynamical simulations. After mimicking the same spatial resolution in the simulated maps than the one used in the observations, we perform a quantitative comparison between the models and the observations using the autocorrelation length. The medium in the different models is driven in time and space and turbulence is charaterized by a particular injection scale $L_{driv}$. Simulations are either isothermal (Models ISO in Tab.~\ref{tab1}) or subject to cooling and heating and might include the effect of self-gravity (models CO in Tab.~\ref{tab1}). Cooling and heating are parametrized with simple functions of the form $\Lambda(\rho,T)=\Lambda_{0}\rho^{\alpha}T^{\beta}$ and $L(\rho)=q\Lambda_{0}\rho^{\epsilon}$, respectively. In the set of simulations discussed in this paper, we essentially vary the parameter which affects the most the efficiency of thermal instability (TI), namely the cooling rate $\Lambda_{0}$. For the isothermal models driven at a Mach number $Ma=1$ which is characteristic of the warm phase of the ISM, the largest over-densities reached in the medium when it reaches dynamical equilibrium are of the order of 2-3 cm$^{-3}$, which are only a few times larger than the average density, $\bar{n}=0.5$ cm$^{-3}$. This contrast to models which include cooling and heating where  TI leads to the formation of condensations with peak densities of a few 100 cm$^{-3}$. The latter condensations usually sit at the intersection of filament and sheet like structures which are separated by low density cavities. In the absence of gravity, the medium settles into a two-phase medium made of cold and dense filaments and warm low density holes and the probability distribution function evolves into a stationary bi-modal form (see Fig.~\ref{fig2}). In the presence of self-gravity, the evolution of the medium is similar to the case without self-gravity until condensations of a few 100 cm$^{-3}$ are formed at $t=\tau_{dyn}$, though these condensations have peak densities that are larger than their counterparts in the non-gravitating case. At a later stage of the evolution, the dense and gravity bound condensations acrrete the surrounding gas to become more massive until they reach the stage of condensation-condensation accretion. The evolution of the autocorrelation length at a given scale is independant from the presence of gravity until this stage of condensation-condensation merger is reached. This is simply due to the fact that, even though the presence of gravity leads to the formation of denser structures, it does not affect the spatial distribution of the condensations in the box in comparison to the non self-gravitating models until the phase of condensation-condensation accretion is reached.

 The autocorrelation length-physical scale relation ($L-L_{cr}$ relation) shows, both in the observations and the simulations, two regimes. For scales smaller than the scale of the energy injection $L_{driv}$, $L_{cr}$ shows a linear increase with $L$ and flattens for scales larger than $L_{driv}$. This is confirmed by a comparison of the observations with synthetic maps derived from simulations driven at different length scales. The $L-L_{cr}$ relation of Ho II implies that energy is injected into the ISM of the galaxy on a scale of $\sim 6$ kpc. This clearly favors a large-scale driving rather than a driving by supernova explosions. The kinetic energy power spectrum drawn by Stanimirovi\'c \& Lazarian (2001) for the Small Magellanic Cloud (SMC) speaks for a large scale driver operating in the SMC as well. The spectrum shows no signs of energy injection up to the largest scale they have considered (4 kpc). The slope of the $L-L_{cr}$ relation seems to be related to the gas physics and the mode of energy decay from $L_{driv}$ down to the smaller scales. Our results show that the slope of the $L-L_{cr}$ relation in the linear regime can be better reproduced when turbulence is coupled to a weak cooling of the gas. We have checked, in the case of model CO09G2 which is the best fitting model to the data, the effect of numerical resolution by performing two additional models at the resolutions of $64^{3}$ and $192^{3}$. We observed the time evolution of the autocorrelation length in the models with different resolutions and found that the three models yield relatively similar values, in particluar models with the resolutions of $128^{3}$ and $192^{3}$, whereas the model with the resolution of $64^{3}$ shows larger fluctuations.

Wada et al. (2000) have shown, in the case of the LMC, that thermal and gravitational instabilities lead to the formation of a network of shell and hole-like structures in the ISM, however, they did not perform a quantitative comparison between their models and the observations in terms of the \ion{H}{i} structure. Our approach can be applied to the ISM in different galaxies in order to 1) constrain the global cooling rate of the galaxy which is connected to the slope of the $L-L_{cr}$ relation in the linear regime and most important, 2) estimate the physical scale on which the energy is injected into the medium by the position of the turning point in the $L-L_{cr}$ curve. In a subsequent paper, using 3D global models of dwarf irregular low star forming galaxies, we plan to investigate the relationship that may exist between the structure function index $\alpha$ ($S(l)=l^{\alpha}$, where $l$ is the physical scale) or the slope of the $L-L_{cr}$ relation in the linear regime for the \ion{H}{i} gas and the global metallicity of the galaxy. Such a relation might exist for dwarf irregular galaxies with small star formation rates and where the gas morphology is likely to be shaped by the interplay of turbulence and the thermodynamical properties of the gas. By building a large library of models including also star formation and feedback, it might also be possible to uncover the relation that exists between the structure function index, the metallicity, and the star formation rate. It is not yet known whether a combination of the metallicity (or a radially dependent metallicity profile) and a radially dependent star formation rate can lead to a unique, non degenerate, large scale morphology of the gas. Establishing such relations between the gas morphology and metallicity, eventually with a radial dependence, would be useful for determining the metallicity of galaxies in a very simple way from their gas morphologies. Interestingly, a large sample of \ion{H}{i} observations for $\sim$ 40 galaxies has been recently obtained with the VLA by F. Walter (Heidelberg) and M. Bureau (Oxford), among them are $\sim$ 10 gas rich dwarf irregular galaxies of different metallicities which can serve as the perfect observational sample to compare our models with.

Regarding the energy requirement of the driver, we have injected into our simulation box $\sim 1.1\times10^{50}$ ergs Myrs$^{-1}$ kpc$^{-3}$. Summing the kinetic energy derived by Puche et al. (1999) (based on kinetic energy input estimates using the model of Chevalier 1974), normalizing that energy input to the age of the oldest hole in Ho II (in time units of Myrs) and dividing by the volume of the galaxy yields a value of $\sim7.3\times10^{48}$ ergs Myrs$^{-1}$ kpc$^{-3}$. This implies that the supernova scenario applied to Ho II underestimates by a factor of $\sim 15$ the required energy for the holes formation. The estimates of Puche et al. (1999) might underestimate the number of supernova explosions in each hole. This would lead to larger remnant stellar clusters than those already expected, but not found, in the \ion{H}{i} holes of Ho II (Rhode et al. 1999). The energy balance argument tends to confirm our conclusions that supernova explosions are not responsible for the formation of all of the observed \ion{H}{i} holes in Ho II.

The natural continuation of this work is to investigate what are the eventual large scale drivers operating in Ho II. In the absence of large scale supernova driving, spiral density waves could constitute a plausible driving candidate. However, Ho II does not show any developed spiral pattern. Another possibility is that Ho II, by infalling to the center of the M81 group and weakly interacting with its neighboring galaxies, might be undergoing a process of ram pressure stripping in the intra-group medium of this group (Bureau and Carignan 2002) and/or tidal stripping by it's closest neighbors. These scenarios could be confirmed by running combined N-Body/SPH simulations of interacting galaxies while including the intergalactic medium gas. Such simulations should be made possible by using new hybrid N-Body/SPH codes like VINE which is currently under development (Wetzstein et al. in preparation). Another possibility to drive turbulence on large scales could be offered by the magneto-rotational instabilities that could take place in the outer parts of the galactic disk of Ho II (Balbus and Hawley 1991; Selwood and Balbus 1999). However, this scenario has not yet been investigated in detail in a galactic context. We should finally mention that the scenario discussed in this paper does not necessarily rule out that \ion{H}{i} holes can also be formed, in other dwarf galaxies, as a result of multiple supernova explosions as discussed by S\'anchez-Salcedo (2002) in the case of the \ion{H}{i} holes observed in the nearby dwarf galaxy IC 2574 (Walter and Brinks 1999).    

\acknowledgements
We would like to thank Katherine Rhode for providing the \ion{H}{i} data of Holmberg II and for useful comments on the first draft. We also would like to thank Fabian Heitsch and Javier Ballesteros-Paredes for motivating discussions at the early stages of this work and are very grateful to the Referee for a careful reading of the manuscript which helped improve it in many aspects. Calculations have been performed on the MPIA's SGI Origin 2000 located at the Rechnenzentrum of the Max-Planck Gesellschaft, Garching. ZEUS-3D was used by courtesy of the Laboratory of Computational Astrophysics at the NCSA. 
 
{}

\clearpage 
\begin{deluxetable}{crrrr}
\footnotesize
\tablecaption{Parameters of the three dimensional simulations \label{tab1}}
\tablewidth{0pt}
\tablehead{
\colhead{Model} & \colhead{$k$} & \colhead{$Ma$} & \colhead{Gravity} & \colhead{$\eta$}}
\startdata

ISO2m1    &2  &1   &no  &0   \\
ISO4m1    &4  &1   &no  &0   \\
ISO2m05   &2  &0.5 &no  &0   \\
ISO2m2    &2  &2   &no  &0   \\
ISO2m3    &2  &3   &no  &0   \\
ISO2m5    &2  &5   &no  &0   \\
CO03G2    &2  &-   &yes &0.3 \\ 
CO03T2    &2  &-   &no  &0.3 \\ 
CO06G2    &2  &-   &yes &0.6 \\
CO09G2    &2  &-   &yes &0.9 \\
\enddata
\end{deluxetable}

\end{document}